\documentclass[8.5pt,twoside,twocolumn,amsmath,amssymb]{article}
\usepackage[super,sort&compress,comma]{natbib}
\usepackage{mhchem}
\usepackage{times,mathptmx}
\usepackage{sectsty}
\usepackage{balance}

\usepackage{graphicx} 
\usepackage{lastpage}
\usepackage[format=plain,singlelinecheck=false,font=small,labelfont=bf,labelsep=space]{caption}
\usepackage{fancyhdr}
\usepackage{dcolumn}
\usepackage{bm}
\usepackage{amsmath}
\usepackage{bbding}
\usepackage{verbatim}
\newcommand{\br}{{\bf r}}
\newcommand{\etal}{{\it et al.}}

\begin{document}

\thispagestyle{plain}
\fancypagestyle{plain}{
\fancyhead[L]{\includegraphics[height=8pt]{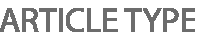}}
\fancyhead[C]{\hspace{-1cm}\includegraphics[height=20pt]{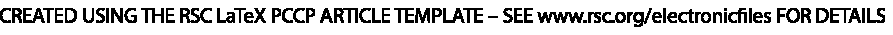}}
\fancyhead[R]{\includegraphics[height=10pt]{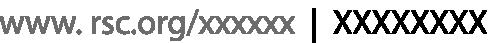}\vspace{-0.2cm}}
\renewcommand{\headrulewidth}{1pt}}
\renewcommand{\thefootnote}{\fnsymbol{footnote}}
\renewcommand\footnoterule{\vspace*{1pt}%
\hrule width 3.4in height 0.4pt \vspace*{5pt}}
\setcounter{secnumdepth}{5}

\makeatletter
\def\subsubsection{\@startsection{subsubsection}{3}{10pt}{-1.25ex plus -1ex minus -.1ex}{0ex plus 0ex}{\normalsize\bf}}
\def\paragraph{\@startsection{paragraph}{4}{10pt}{-1.25ex plus -1ex minus -.1ex}{0ex plus 0ex}{\normalsize\textit}}
\renewcommand\@biblabel[1]{#1}
\renewcommand\@makefntext[1]%
{\noindent\makebox[0pt][r]{\@thefnmark\,}#1}
\makeatother
\renewcommand{\figurename}{\small{Fig.}~}
\sectionfont{\large}
\subsectionfont{\normalsize}

\fancyfoot{}
\fancyfoot[LO,RE]{\vspace{-7pt}\includegraphics[height=9pt]{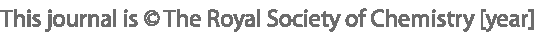}}
\fancyfoot[CO]{\vspace{-7.2pt}\hspace{12.2cm}\includegraphics{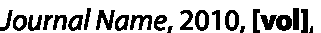}}
\fancyfoot[CE]{\vspace{-7.5pt}\hspace{-13.5cm}\includegraphics{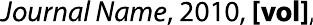}}
\fancyfoot[RO]{\footnotesize{\sffamily{1--\pageref{LastPage} ~\textbar  \hspace{2pt}\thepage}}}
\fancyfoot[LE]{\footnotesize{\sffamily{\thepage~\textbar\hspace{3.45cm} 1--\pageref{LastPage}}}}
\fancyhead{}
\renewcommand{\headrulewidth}{1pt}
\renewcommand{\footrulewidth}{1pt}
\setlength{\arrayrulewidth}{1pt}
\setlength{\columnsep}{6.5mm}
\setlength\bibsep{1pt}

\twocolumn[
  \begin{@twocolumnfalse}
\noindent\LARGE{\textbf{Ion Density Deviations in Polyelectrolyte Microcapsules: \\
 Influence on Biosensors}}
\vspace{0.6cm}

\noindent\large{\textbf{Qiyun Tang\textit{$^{a}$} and Alan R. Denton\textit{$^{\ast}$\textit{$^{a}$}}
}}\vspace{0.5cm}

\noindent\textit{\small{\textbf{Received 24th June 2014, Accepted 22nd August 2014}}}

\noindent \textbf{\small{DOI: 10.1039/c4cp02773f}}
\vspace{0.6cm}

\noindent \normalsize{Polyelectrolyte microcapsules loaded with fluorescent dyes have been
proposed as biosensors to monitor local pH and ionic strength for
diagnostic purposes. In the case of charged microcapsules, however,
the local electric field can cause deviations of ion densities
inside the cavities, potentially resulting in misdiagnosis of some
diseases.  Using nonlinear Poisson-Boltzmann theory, we
systematically investigate these deviations induced by charged
microcapsules. Our results show that the microcapsule charge
density, as well as the capsule and salt concentrations, contribute
to deviations of local ion concentrations and pH.  Our findings are
relevant for applications of polyelectrolyte microcapsules with
encapsulated ion-sensitive dyes as biosensors.}
\vspace{0.5cm}
 \end{@twocolumnfalse}
  ]

\footnotetext{\textit{$^{a}$~Department of Physics, North Dakota State University, Fargo, ND 58108-6050, USA.
E-mail: alan.denton@ndsu.edu}}

\section{Introduction}

Polyelectrolyte (PE) microcapsules, polymeric particles whose hollow
cavities can be loaded with dye molecules~\cite{Borisov_2010_AFR},
have attracted great attention in the past decade due to their
potential biomedical applications~\cite{delMercato_2010_Nanoscale},
such as therapeutic drug
delivery~\cite{DeKoker_2012_CSR,Koker_2011_ADDR} and
diagnostics~\cite{DeGeest_2009_Softmatter,Weijun_2012_CSR,Adamczak_2012_CSB,
Peteiro_2009_Nanomedicine}. Among the various experimental methods
developed to fabricate microcapsules, one of the most attractive is
layer-by-layer (LBL) assembly, due to its precise control of the
capsule size, thickness, shape, and
functions~\cite{Caruso_1998_Science}. The LBL technique uses diverse
driving forces, such as electrostatic
interactions~\cite{DeGeest_2009_Softmatter}, van der Waals
interactions~\cite{Kida_2006_Angew}, hydrogen
bonding~\cite{Such_2011_CSR,Kharlampieva_2006_PolymRev}, guest-host
interactions~\cite{Wang_2008_CM}, covalent
bonding~\cite{Zhang_2003_MA,Duan_2007_BBRC}, and base-pair
interactions~\cite{Johnston_2005_NanoLett}, to deposit the
multilayer films onto colloidal and nanoparticle templates, followed
by the removal of sacrificial templates. When their cavities are
loaded with therapeutic drugs, fluorescent dyes, or chemical
reactants, such microcapsules can function as drug delivery
vehicles~\cite{RiveraGil_2009_NL,Koker_2011_ADDR}, precise optical
ratiometric
biosensors~\cite{Shen_2011_MA,Reibetanz_2011_JBS,Kazakova_2011_PCCP,Kreft_2007_JMC,
McShane_2010_JMC}, or
bioreactors~\cite{Staedler_2009_Nano,Shchukin_2004_AM}.

Recently, much interest has focused on applications of PE
microcapsules loaded with ion-sensitive dyes as biosensors to
monitor local ion concentrations, such as pH, in cellular
environments~\cite{Kuwana_2004_BP,Reibetanz_2010_BioM,
deMercato_2011_Small,Kreft_2007_JMC,DeGeest_2009_Softmatter,delMercato_2011_ACSNano,
Xiaoxue_2014_JCIS,
Kazakova_2013_ABC,Sun_2012_BC,Dorleta_2012_CM,RiveraGil_2012_Small,
Lee_2011_Softmatter}. For example, Kreft \etal~\cite{Kreft_2007_JMC}
introduced PE microcapsules loaded with pH-sensitive
SNARF-1 dye molecules to monitor the local pH within the alkaline
cell medium and the acidic endosomal/lysosomal compartments.
Recently, triple-labeled PE microcapsules loaded with two
pH-sensitive dyes and one pH-insensitive dye, were also fabricated
to measure the local pH in real time in living
cells~\cite{Xiaoxue_2014_JCIS}. Del Mercato
\etal~\cite{delMercato_2011_ACSNano} demonstrated that double-wall
barcoded sensor capsules can be used for multiplexed detection of
protons and sodium and potassium ions in parallel. These
capsule-based sensor systems offer advantages in practical
applications. First, the dye molecules are encapsulated in small
cavities at high local concentrations, leading to high resolution
optical images. Second, the dye molecules can be encapsulated in
biocompatible PE microcapsules~\cite{Patel_2013_AFM} to circumvent
the toxicity in cellular environments and detect the local ion
concentration in real time. Third, sensing and labeling dyes can be
separately loaded in the cavities and walls of the microcapsules,
providing a promising procedure for multiplexed
sensing~\cite{delMercato_2011_ACSNano}. These microcapsule-based
chemical sensors have potential biomedical applications in the
diagnosis of certain
diseases~\cite{DeGeest_2009_Softmatter,Weijun_2012_CSR,Adamczak_2012_CSB,
Peteiro_2009_Nanomedicine}.

In the LBL technique, microcapsules are often formed by exploiting electrostatic
interactions to alternately adsorb PE layers onto oppositely charged templates.
The PE shells, which become charged by dissociating counterions into aqueous
solution, generate an electric field that influences the ion distributions
(such as local pH) near the microcapsule shells.  Previous work has shown that
the ionic strength variations between bulk and charged surface regions may be
significant~\cite{Janata_1987_AC,Janata_1992_AC,Bostrom_2002_La,Zhang_2011_CPC}.
For example, Janata~\cite{Janata_1987_AC,Janata_1992_AC} pointed out that
bulk-surface interactions should be considered for the pH shift in optical sensors.
Bostrom \etal~\cite{Bostrom_2002_La} showed that ion-specific dispersion
potentials near biological flat membranes could induce ion and pH gradients.
Zhang \etal~\cite{Zhang_2011_CPC} experimentally demonstrated that surface charges
can contribute significantly to local ion concentrations and sensor read-out.
This phenomenon influences, in turn, the measured concentrations of ions in
microcapsule cavities.  Understanding these deviations is of great significance
in biomedical applications, for example, to avoid misdiagnosis of diseases, such as
early-stage cancer~\cite{Weidgans_2006_Science,RiveraGil_2008_ACSNano,Xie_2011_ACR}.
Previous experiments~\cite{Sukhorukov_1999_JPCB,David_2005_JCIM} and
theoretical models~\cite{Sukhorukov_1999_JPCB,David_2007_ACS}
have demonstrated that the pH inside microcapsules can differ from that outside.
This difference was attributed mainly to semipermeability of the capsules,
which impedes diffusion of one species of small ion across the capsule wall 
and generates Donnan equilibrium. 
To the best of our knowledge, however, there is still no theoretical work to
quantitatively investigate the deviations in ion concentration induced by
charged microcapsules.

The goal of this paper is to analyze the impact of the electric field of ionized
PE shells on the performance of microcapsules as biosensors in aqueous solutions.
By employing nonlinear Poisson-Boltzmann theory, we systematically calculate the
deviation of ion density inside microcapsule cavities induced by charged shells.
We find that the redistribution of ions inside the cavities of negatively charged
capsules shifts the pH, as would be measured by encapsulated fluorescent dyes,
towards lower (more acidic) values.  The magnitude of the shift strongly depends on
the charge density of the microcapsules, and on the capsule and salt concentrations.
Our work differs from previous reports of pH
deviations~\cite{Sukhorukov_1999_JPCB,David_2005_JCIM,David_2007_ACS}
in that, in our model, the capsule wall is permeable to all ion species, 
and the predicted deviations of ion concentration and pH are not related to
semipermeability of the capsules.
Our results demonstrate that electric charge can influence the ability of ionic
microcapsules to function as biosensors, which should stimulate experiments
and simulations.

The remainder of the paper is organized as follows.  In the next section,
we describe our model and theoretical methods.  In Sec.~\ref{results},
we present results for the influence of charge density, and of capsule and salt
concentrations, on the variation of ion distributions between the cavity and
bulk regions.  We then discuss advantages and limitations of our methods and
prospects for future studies.  Finally, in Sec.~\ref{conclusions}, we summarize
and emphasize implications for practical applications.

\begin{figure}[h]
 \includegraphics[width=8cm,bb=0 0 1010 404]{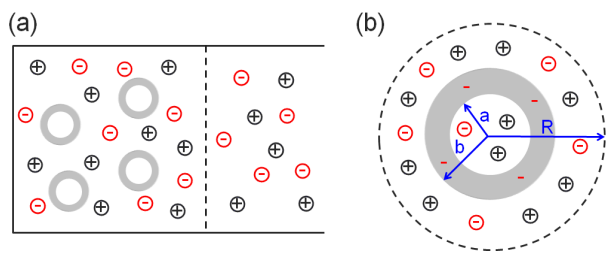}
  \caption{(a) Primitive model: an aqueous solution of polyelectrolyte microcapsules
  (shaded rings) and positive and negative microions dispersed in a uniform dielectric
  (water).  (b) Cell model: a single microcapsule, modeled as a uniformly
  charged shell (shaded ring) of valence $Z$ and inner and outer radii $a$ and $b$,
  centered in a spherical cell of radius $R$. }\label{fig1}
\end{figure}

\section{Models and Methods}\label{models}
We consider a solution of microcapsules, consisting of spherical shells of
cross-linked PE chains.  In a polar solvent (here water), the shells acquire
charge through dissociation of counterions from the polymer backbones.
The shells are permeable to water and small ions and enclose cavities that
are filled with an aqueous solution, whose ionic strength can differ from
that of the bulk solution.
In Donnan equilibrium, the microcapsules are confined to a fixed volume,
while the microions (counterions, salt ions) and solvent can freely exchange
with a salt reservoir, e.g., via a semi-permeable membrane [Fig.~\ref{fig1}(a)].

\begin{figure}[h]
 \includegraphics[width=8cm,bb=0 0 792 1109]{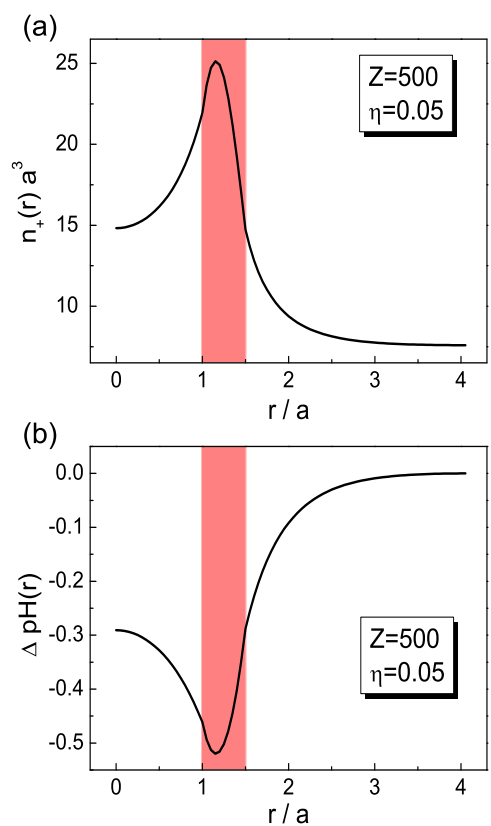}
  \caption{(a) Reduced cation number density and (b) deviation of local pH from
  its bulk value (at cell edge) vs.~radial distance from center of a PE
  microcapsule of inner shell radius $a=50$ nm, outer radius $b=75$ nm,
  valence $Z=500$, and dielectric constant ratio $\chi=0.5$, in an aqueous solution
  of microcapsule volume fraction $\eta=0.05$ and reservoir salt concentration $n_0=0.1$ mM.
  Shaded bar indicates shell region.
  }\label{fig2}
\end{figure}

We start from the primitive model~\cite{Marcus_1955_JCP}, which
reduces the solvent to a dielectric continuum of dielectric constant
$\epsilon$, and further idealize the dense network of PE chains and
water within a microcapsule shell as a uniform medium with
dielectric constant $\epsilon_{\rm shell}$, fixed charge $Ze$, and
uniform charge number density, $n_f(r)=Z/V_{\rm shell}$ $(a<r<b)$,
confined to a volume $V_{\rm shell}=(4\pi/3)(b^3-a^3)$, between the
inner radius $a$ and outer radius $b$ [Fig.~\ref{fig1}(b)]. We
assume that the water within the cavity ($r<a$) has a dielectric
constant equal to that in bulk.  In the PE shell, however, we assume
$\epsilon_{\rm shell}<\epsilon$, since PE microgels are known to
have dielectric constants lower than that of bulk
water~\cite{Parthasarathy_1996_MSE,Mohanty_2012_SM}. Finally, we
model the microions simply as point charges. Although neglecting
charge discreteness and ion-specific effects due to hydration, this
coarse-grained model is constructed to capture the essential
physical features of real PE microcapsules.

The substantial asymmetry in size and charge between polyions and
microions motivates the cell model~\cite{Hakan_1982_JCP}.  For
spherical polyions, such as microcapsules, the cell model represents
a bulk solution by a spherical cell of radius $R$, centered on a
single polyion.  The cell contains a neutralizing number of
counterions and coions in a volume determined by the polyion volume
fraction $\eta=(b/R)^3$ [Fig. \ref{fig1}(b)].  For simplicity, we
assume here monovalent microions, whose correlations with one
another are sufficiently weak to justify a mean-field approach, in
particular, a Poisson-Boltzmann theory.

The Poisson-Boltzmann (PB) theory of polyelectrolyte solutions
combines the exact Poisson equation with the Boltzmann approximation
for the ion density
distribution~\cite{Marcus_1955_JCP,Hakan_1982_JCP,Denton_2003_PRE,Denton_2010_JPCM}.
In the primitive model, in which the solvent has a uniform
dielectric constant $\epsilon$, the Poisson equation,
\begin{equation}
\begin{split}
\nabla^2\phi(\br)=-\frac{4\pi}{\epsilon}\rho(\br)~,
\end{split}
\label{Poisson1}
\end{equation}
relates the electrostatic potential $\phi(\br)$ at position $\br$ to the
local charge density, $\rho(\br)=e[n_+(\br)-n_-(\br)-n_f(\br)]$, which
includes the number densities of both mobile microions, $n_{\pm}(\br)$,
and fixed ions on the PE chains, $n_f(\br)$.
Equation~(\ref{Poisson1}) can also be expressed as
\begin{equation}
\begin{split}
\nabla^2\psi(\br)=-4\pi \lambda_B[n_+(\br)-n_-(\br)-n_f(\br)]
\end{split}\label{Poisson2}
\end{equation}
by defining the reduced potential, $\psi\equiv e\phi/k_BT$, and the Bjerrum length,
$\lambda_B\equiv e^2/\epsilon k_BT$, as the distance at which two elementary
charges $e$ interact with the typical thermal energy $k_BT$
at absolute temperature $T$.
In Donnan equilibrium, the Boltzmann approximation for the microion number
density distributions is
\begin{equation}
n_{\pm}(\br)=n_0\exp[\mp\psi(\br)]~,
\label{Boltzmann}
\end{equation}
where $n_0$ is the average number density of salt ion pairs in the reservoir.
This mean-field approximation neglects short-range ion-ion correlations.

Combining the Poisson equation for the electrostatic potential [Eq.~(\ref{Poisson2})]
with the Boltzmann approximation for the microion density distributions
[Eq.~(\ref{Boltzmann})] yields the PB equation,
\begin{equation}
\nabla^2\psi(\br)=\kappa^2\sinh\psi(\br)+4\pi\lambda_B n_f(\br)~,
\label{PBeqn1}
\end{equation}
where $\kappa=\sqrt{8\pi\lambda_Bn_0}$ is the screening constant in the
salt reservoir.
Implementation of the PB theory is facilitated by adopting the cell model,
wherein solution of Eq.~(\ref{PBeqn1}) is greatly eased by geometric symmetry
and neglect of polyion-polyion correlations.  In the spherical cell model,
the PB equation simplifies to
\begin{equation}
\psi''(r)+\frac{2}{r}\psi'(r)=\begin{cases} \kappa^2\sinh\psi(r)~,
&0<r<a~,\\
{\displaystyle \frac{\kappa^2}{\chi}\sinh\psi(r)
+\frac{3Z\lambda_B/\chi}{b^3-a^3}}~,
&a<r<b~,\\
\kappa^2\sinh\psi(r)~,
&b<r<R~,
\end{cases}\label{PBeqn2}
\end{equation}
where $r$ is the radial distance from the center of the cell and
$\chi=\epsilon_{\rm shell}/\epsilon<1$ is the ratio of the dielectric constant
in the microcapsule shell to that in the bulk solvent.

The boundary conditions on Eq.~(\ref{PBeqn2}) are clear.  First, the electrostatic
potential is continuous at the inner and outer boundaries of the microcapsule shell.
Labelling the solutions in the three regions as
$\psi_{\rm in}(r)$ ($0<r<a$), $\psi_{\rm shell}(r)$ ($a<r<b$), and
$\psi_{\rm out}(r)$ ($b<r<R$), we have
\begin{equation}
\psi_{\rm in}(a)=\psi_{\rm shell}(a)~, \hspace{1cm}
\psi_{\rm shell}(b)=\psi_{\rm out}(b)~.
\label{boundP}
\end{equation}
Second, spherical symmetry requires that the electric field vanish at
the center of the cell, while Gauss's law and electroneutrality require
that the electric field vanish on the cell boundary:
\begin{equation}
\psi'_{\rm in}(0)=0~, \hspace{1cm} \psi'_{\rm out}(R)=0~.
\label{boundE1}
\end{equation}
Finally, continuity of the electric displacement on the inner and outer
shell boundaries requires
\begin{equation}
\psi'_{\rm in}(a)=\chi\psi'_{\rm shell}(a)~,
\hspace{1cm}
\chi\psi'_{\rm shell}(b)=\psi'_{\rm out}(b)~.
\label{boundE2}
\end{equation}
We calculate the equilibrium microion distributions within the spherical cell
by numerically solving the PB equation [Eq.~(\ref{PBeqn2})], along with the
boundary conditions [Eqs.~(\ref{boundP})-(\ref{boundE2})], in the three
radial regions (inside the cavity, in the shell, and outside the capsule),
matching the solutions at the inner and outer shell boundaries
using a two-dimensional root-finding algorithm~\cite{Numerical_Recipes_2007}.

As an illustration of our method, Fig.~\ref{fig2}(a) shows the cation
number density in the vicinity of a negatively charged PE microcapsule
of inner shell radius $a=50$ nm, outer shell radius $b=75$ nm, and
valence $Z=500$.  The PE microcapsules are dispersed in room-temperature
water ($\lambda_B=0.72$ nm), with reservoir salt concentration $n_0=$0.1 mM,
at a volume fraction $\eta=0.05$ -- corresponding to a cell radius
$R=\eta^{-1/3}b\simeq 4.07~a$.  At such dilution, the microcapsules
are sufficiently separated that the ion distributions within different
cavities are essentially independent.  The dielectric constant ratio is set
as $\chi=0.5$, which is consistent with hydrated PNIPAM microgels,
whose dielectric constant ranges from 63 at 15$^{\circ}$C to 17 at
40$^{\circ}$C~\cite{Parthasarathy_1996_MSE,Mohanty_2012_SM}.
Figure~\ref{fig2}(a) demonstrates that the ion density inside the
capsule exceeds the bulk value (at the cell boundary), confirming that
the charged shell indeed redistributes the ion density
near the microcapsule. Note the discontinuous slopes in Fig.~\ref{fig2}
at the shell boundaries ($r=a, b$), which simply reflect the
jump in dielectric constant between the shell and water.

Many recent experiments have focused on measurements of local pH
using fluorescent dyes encapsulated in PE microcapsules. Within
our theoretical framework, the local pH inside the PB cell is
normally determined via ${\rm pH}(r)=-\log([{\rm H}^+](r))$, where
$[{\rm H}^+](r)$ denotes the hydronium ion concentration (in M) at
radial distance $r$ from the cell center. The local deviation of pH from
its bulk value is then obtained via $\Delta{\rm pH}(r)=-\log([{\rm H}^+](r)/
[{\rm H}^+]_{\rm bulk})$, where $[{\rm H}^+]_{\rm bulk}$ is
the bulk hydronium concentration. In an alkaline environment,
the pH can be defined, alternatively, from the concentration of
hydroxide (${\rm OH}^-$) ions, giving rise to the deviation
$\Delta{\rm pH}(r)=\log([{\rm OH}^-](r)/[{\rm OH}^-]_{\rm bulk})$.

Because the mean-field PB theory ignores ion-specific effects, it cannot
distinguish hydronium ions from other monovalent cations, such as
Na$^+$ or K$^+$. Nevertheless, in the presence of an inhomogeneous
electric field generated here by a charged microcapsule shell, the
local deviation of hydronium concentration from its bulk value
should be proportional to the local deviation of the concentration
of all monovalent cations. Thus, we approximate the local deviation
of pH from its bulk value by
\begin{equation}
\Delta{\rm pH}(r)\simeq \begin{cases} -\log[n_{+}(r)/n_+(R)]
&\text{\hspace{0.15cm}(acidic)}\\[1ex]
\log[n_{-}(r)/n_-(R)]&\text{(alkaline)}
\end{cases}
\label{delta-pH}
\end{equation}
where $n_+(R)$ and $n_-(R)$ are the cell model approximations for the
bulk cation and anion densities, respectively.
In the Boltzmann approximation [Eq.~(\ref{Boltzmann})], pH deviation
is proportional to the electrostatic potential deviation:
\begin{equation}
\Delta{\rm pH}(r)=(\log e)[\psi(r)-\psi(R)]~,
\label{delta-pH2}
\end{equation}
illustrating the direct connection between pH deviation and the electrostatic
influence of the microcapsule, and providing a practical formula for calculations.
Figure~\ref{fig2}(b) shows the deviation of local pH based on the cation distribution in
Fig.~\ref{fig2}(a). For the chosen parameters, the local pH inside the
negatively charged shell is lower than the bulk value within a range
from 0.3 to 0.5. Also of experimental relevance is the average
pH deviation, defined as an average over the volume of the cavity:
\begin{equation}
\langle\Delta{\rm pH}\rangle\equiv\frac{3}{a^3}\int_0^a dr\, r^2
\Delta{\rm pH}(r)~. \label{pHav}
\end{equation}
In the next section, we explore the dependence of ion distributions
and pH deviation on the valence, size, and concentration of
negatively charged microcapsules.

\section{Results and Discussion}\label{results}

\begin{figure}[h]
 \includegraphics[width=8cm,bb=0 0 792 1109]{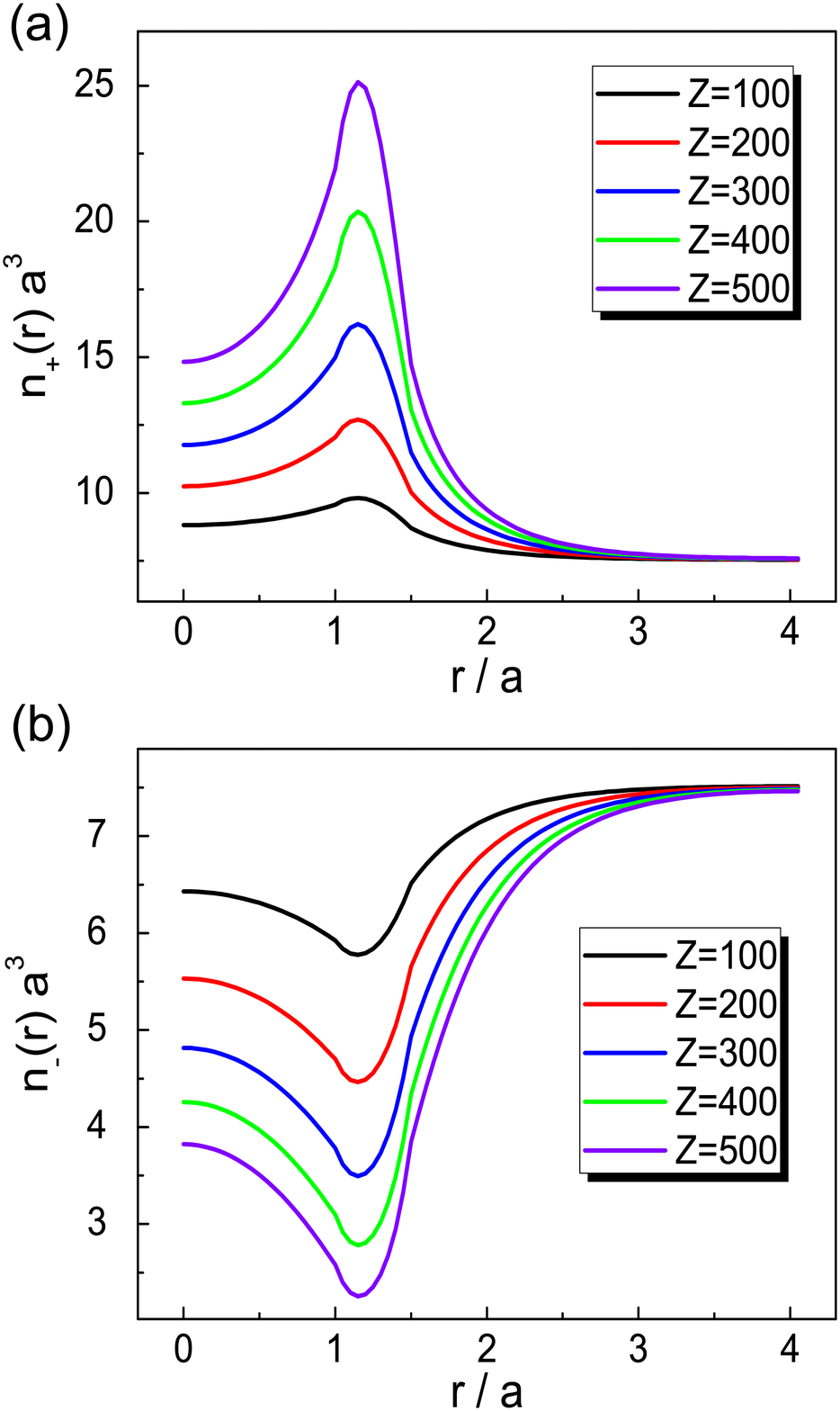}
  \caption{Number densities of (a) cations and (b) anions inside cell
  for microcapsule valences $Z=100$, 200, 300, 400, and 500 [bottom to top
  in (a), top to bottom in (b)].
  (Other parameters are as in Fig.~\ref{fig2}.)
  }\label{fig3}
\end{figure}

\begin{figure}[h]
 \includegraphics[width=8cm,bb=0 0 792 1109]{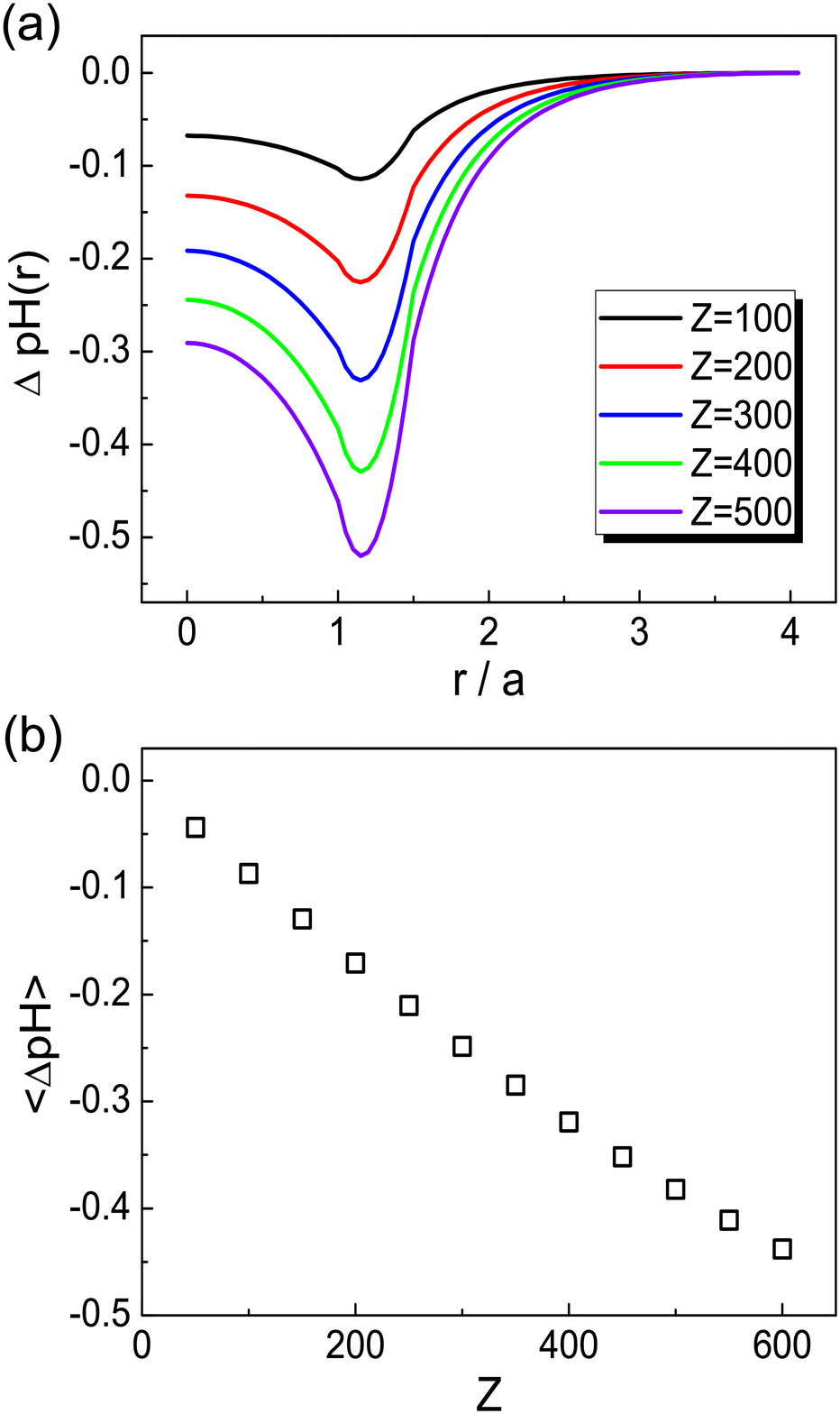}
  \caption{(a) Deviation of local pH from its bulk value vs.~radial distance
  from microcapsule center for valences $Z=100$, 200, 300, 400, and 500 (top to bottom).
  (b) Average pH deviation inside cavity vs.~valence.
  (See Fig.~\ref{fig2} for other parameters.)
  }\label{fig4}
\end{figure}

\begin{figure}[h]
 \includegraphics[width=8cm,bb=0 0 792 1109]{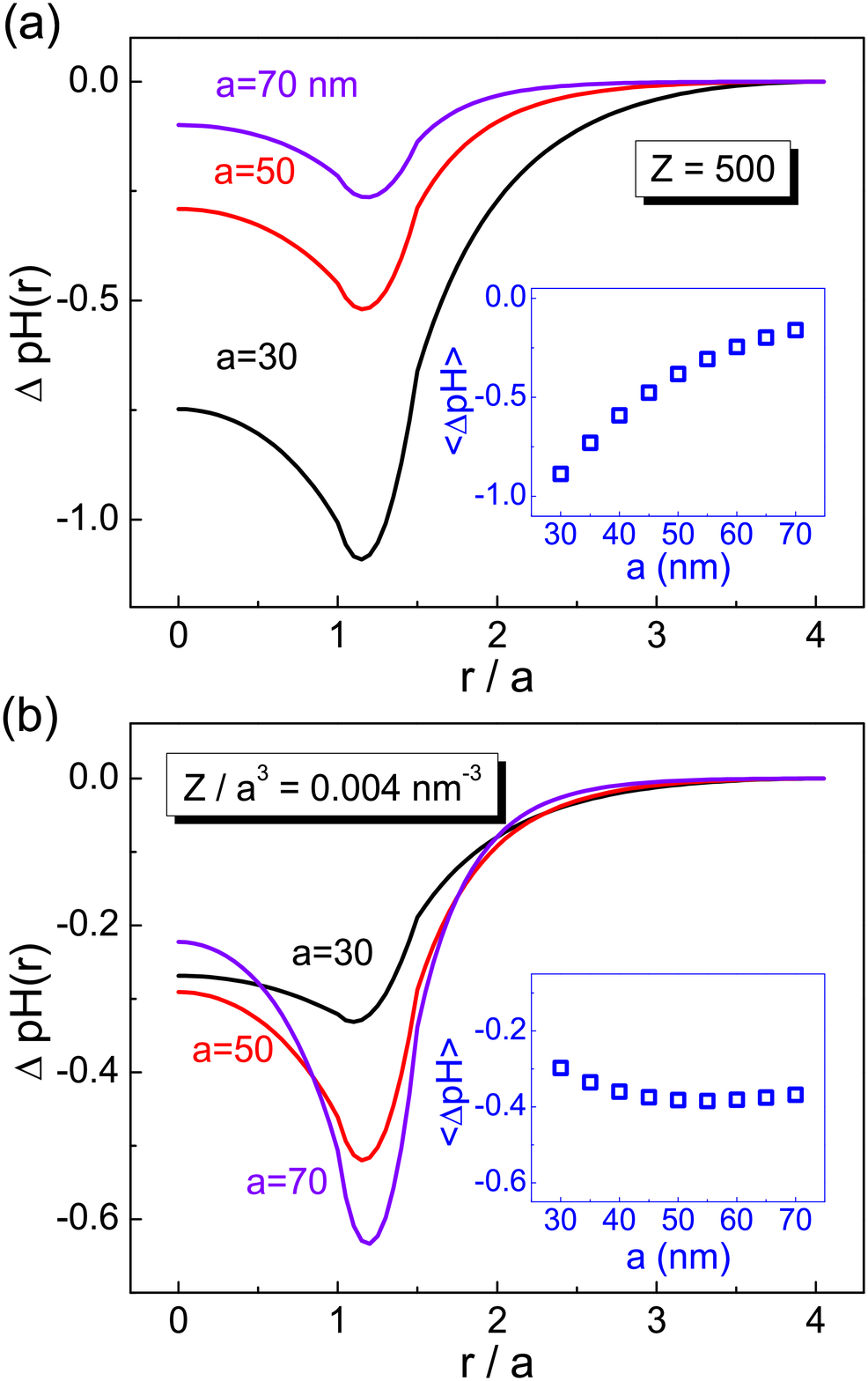}
  \caption{Deviation of local pH vs.~radial distance from microcapsule center for
  cavity inner radius $a=30$, 50, 70 nm (outer radius $b=1.5a=$45, 75, 105 nm) at
  (a) fixed capsule valence $Z=500$ and (b) fixed charge density $Z/a^3=0.004$ nm$^{-3}$.
  Insets show variation of average pH deviation inside cavity with cavity radius.
  (Other parameters are as in Fig.~\ref{fig2}.)
  }\label{fig5}
\end{figure}

\subsection{Influence of Charge Density on pH Deviations 
inside Microcapsule Cavities}

In this section, we systematically investigate the influence of a negatively
charged microcapsule on ion density distributions within the PB cell model.
In experiments, the microcapsule valence can be tuned by varying the shell
composition, as well as bulk solution conditions (pH and salt concentration),
and even temperature.
Figure~\ref{fig3} shows the dependence of the ion distributions on the
microcapsule valence, demonstrating that the cation density is higher than
the bulk value throughout the cell, reaches a maximum within the shell,
and increases with valence, from $Z=100$ to $Z=500$.
Furthermore, with increasing microcapsule valence, the maximum increases,
while the bulk cation density remains nearly constant, implying that the
deviations of cation density induced by the charged shell are
strongly determined by the capsule valence.
Conversely, the anion density is lower than the bulk value, reaches a minimum
within the shell, and decreases steadily with increasing valence.
These deviations result from the attraction of cations (repulsion of anions)
by the negatively charged shell.

In an acidic environment, the cation distribution induced by the charged
capsule correlates with the deviation of pH from its bulk value, according to
Eq.~(\ref{delta-pH}).  As an example, Fig.~\ref{fig4} shows the distributions
of local and average pH deviations over a range of microcapsule valences.
With increasing valence, the pH inside the microcapsules falls increasingly
below the bulk pH value, while the average pH deviation inside the capsule
cavities increases in magnitude.  At the highest valence considered here,
the pH deviation from its bulk value in the cavity can approach $-0.5$.
A shift of this magnitude may be significant for proposed applications of
PE microcapsules as pH sensors~\cite{Kreft_2007_JMC}, where encapsulated
dye molecules fluoresce at frequencies that depend on the local pH.
In an alkaline environment, essentially the same pH deviation is obtained
from Eq.~(\ref{delta-pH}) when calculated using the anion density distribution.
It is important to remember, however, that the valence of a PE microcapsule
in solution may itself depend self-consistently on the local pH, since local pH
can affect dissociation-association equilibrium and thus the degree of
ionization of the PE chains making up the shell.

\begin{figure}[h!]
 \includegraphics[width=8cm,bb=0 0 792 1664]{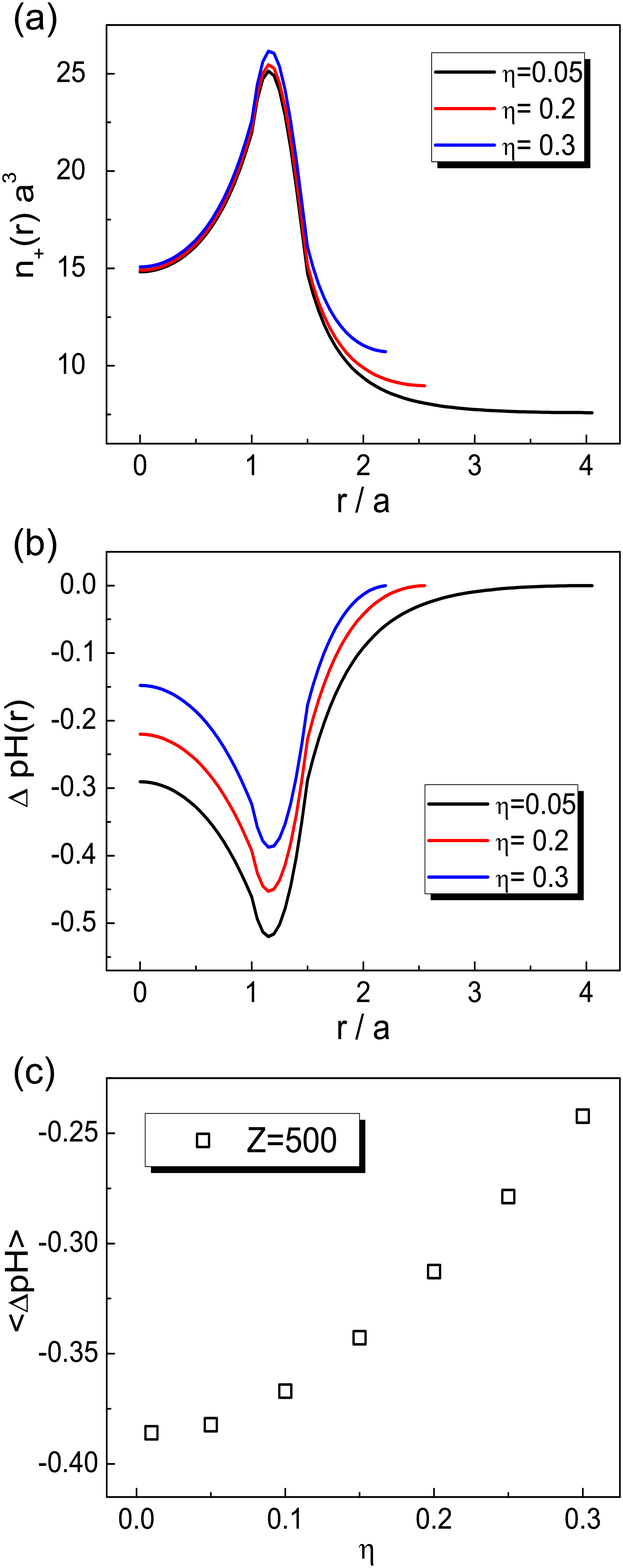}
  \caption{(a) Cation number density and (b) local pH deviation for microcapsules
  of valence $Z=500$ and volume fraction $\eta=$0.05, 0.2, and 0.3 (bottom to top).
  (c) Average pH deviation in cavity vs.~$\eta$.
  (See Fig.~\ref{fig2} for other parameters.)
  }\label{fig6}
\end{figure}
\begin{figure}[h!]
 \includegraphics[width=8cm,bb=0 0 792 1109]{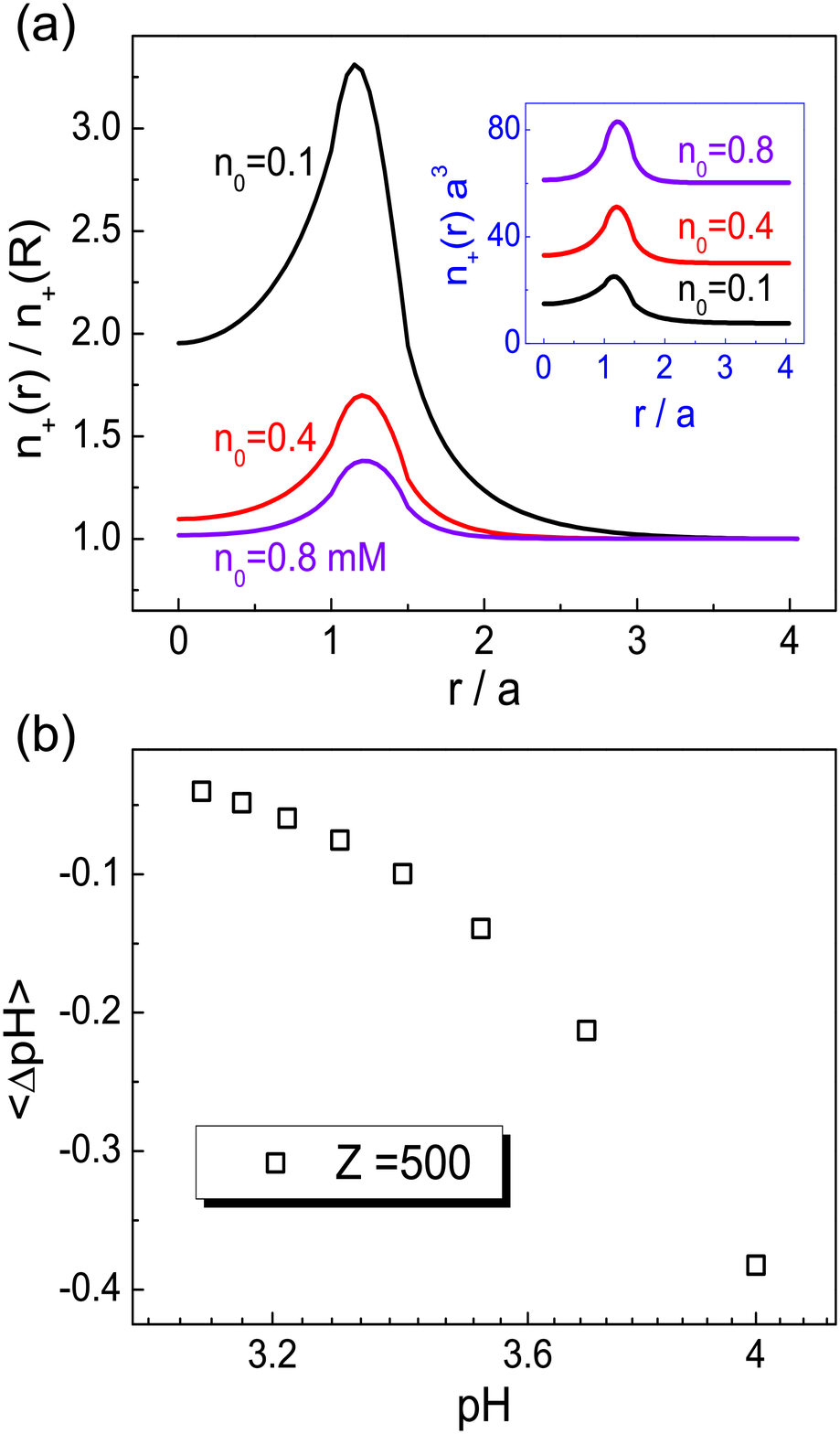}
  \caption{(a) Local cation density, as a fraction of bulk density,
  at reservoir salt concentration $n_0=0.1$, 0.4, and 0.8 mM.
  Inset shows the corresponding cation number densities.
  (b) Average pH deviation vs.~pH at $Z=500$ and $a=50$ nm.
  (See Fig.~\ref{fig2} for other parameters.)
  }\label{fig7}
\end{figure}

Next, we investigate the influence of microcapsule size on induced
pH deviations. Although many recent experiments involve capsules
that are a few microns in
size~\cite{Kreft_2007_JMC,Xiaoxue_2014_JCIS,delMercato_2011_ACSNano},
here we consider considerably smaller capsules, tens of nanometers
in radius, which is the current limit of our numerical methods.  As
shown in Fig.~\ref{fig5}(a), for fixed valence, as the cavity inner
radius increases, the local and average pH deviations decrease in
magnitude.  This trend is to be expected, since for the same charge,
the electric field inside of a larger capsule is weaker, resulting
in less redistribution of ions.  However, if the capsule size and
valence vary together, at fixed valence {\it density}, the pH
deviation is relatively insensitive to the capsule size (see
Fig.~\ref{fig5}(b)).  It is worth noting that the pH deviations
exhibited in Fig.~\ref{fig4} also reflect the dependence on charge
density (with fixed capsule size). In experiments, charge densities
are determined by both the density of the PE network and the degree
of PE dissociation in solution.  Extrapolation to larger capsules
suggests that pH deviations inside the cavities of strongly
dissociated PE capsules of the size studied in recent
experiments~\cite{Kreft_2007_JMC,Xiaoxue_2014_JCIS,
delMercato_2011_ACSNano} may be significant.

From the above considerations, we conclude that charge density of PE
shells is a major determiner of deviations of ion density and pH
induced by charged microcapsules. Next, we show that even when the
charge density remains constant, other factors can still influence
ion concentrations inside of charged microcapsules.

\subsection{Influence of Microcapsule and Salt Concentrations on pH Deviations}

Among other factors that can affect ion density and pH deviations in microcapsule
cavities, the first that we consider is the microcapsule volume fraction $\eta$,
whose influence is illustrated in Fig.~\ref{fig6}.  For fixed shell charge density
($Z=500$, $a=50$ nm), Fig.~\ref{fig6}(a) shows that the cation density inside the cavity
remains essentially unchanged with increasing $\eta$ (i.e., decreasing cell radius $R$),
while outside the microcapsule, the bulk cation concentration $n_+(R)$ steadily increases.
Correspondingly, the pH deviation inside the cavity decreases with increasing $\eta$
[Figs.~\ref{fig6}(b) and \ref{fig6}(c)].
It follows that an inhomogeneous distribution of PE microcapsules can yield
a spatial variation of pH deviation and that, in concentrated solutions,
pH values measured from fluorescence of encapsulated dye
molecules should be adjusted for concentration dependence.

Another factor that can affect deviations of ion densities in capsule cavities
from their bulk values is the reservoir salt concentration $n_0$. As illustrated
in Fig.~\ref{fig7}, for fixed capsule charge density ($Z=500$, $a=50$ nm),
the deviation of cation density (and, therefore, pH) diminishes with increasing
salt concentration.
In the extreme case of a solution that contains predominantly H$^+$ cations,
the reservoir salt concentration $n_0$ is directly related to the reservoir pH
value via pH$=-\log n_0$.  For such a deionized system, Fig.~\ref{fig7}(b) shows
that the average pH deviation inside the cavity induced by the charged shell
is then strongly dependent on the bulk pH.
This dependence indicates that the local pH measured by capsule-based sensors
in low-pH acidic environments can deviate less than in acidic environments
with higher pH.
Therefore, in practical applications, the calibration of measured
local pH to the actual values in cellular environments should also
take into account the gradient of the pH deviation induced by the
absolute local pH.

It should be mentioned that ion concentrations may depend also on temperature.
This dependence is somewhat complicated, however, by the sensitivity of counterion
dissociation and Donnan equilibrium to thermal fluctuations.  Therefore, 
in this work, we simply assume fixed (room) temperature.

\subsection{Discussion}
We have demonstrated here that negatively charged PE
microcapsules redistribute ion concentrations, resulting in
accumulations of H$^+$ ions in their cavities and corresponding
shifts of local pH to values lower than in bulk.  The magnitude of
the pH shift depends strongly on the microcapsule charge density,
as well as on the capsule and salt ion concentrations. These results
confirm that the ion concentrations measured by dye molecules
encapsulated in charged PE microcapsule cavities can deviate from
their bulk values, potentially leading to misdiagnosis of
diseases~\cite{Weidgans_2006_Science,RiveraGil_2008_ACSNano,Xie_2011_ACR}.
In this section, we discuss some limitations of Poisson-Boltzmann
theory and several open questions for further investigation.

First, the PB theory implemented here
cannot distinguish distinct monovalent ions.  In this paper, we assume that
density deviations induced by charged PE microcapsules are the same
for all monovalent cations and choose H$^+$ as an example to
illustrate the deviation of pH in the cavities of charged
microcapsules. Our results can be easily extended to the detection
of other monovalent ions, such as Na$^+$ and K$^+$.  Previous studies
have shown, however, that ionic strength can influence the local pH
obtained from optical pH
sensors~\cite{Janata_1987_AC,Janata_1992_AC,Weidgans_2004_Analyst}.
Thus, it is still not clear how specific interactions between
different monovalent cations may influence deviations of ion
concentrations and pH induced by charged microcapsules. Experiments
and further modeling are required to address this issue.

A second issue deserving further discussion is the magnitude of the
local pH deviation induced by negatively charged microcapsules. For
the parameters considered in this paper, the pH shift is typically
within the range 0.1 to 0.6.  Although higher shell charge densities
can induce shifts exceeding 1.0, the energy barrier across the shell
in this case is larger than 2 $k_BT$, which kinetically impedes ions
from penetrating the capsule by thermal diffusion.  The typical
deviation magnitude of 0.5 observed in this paper is within the
error bar of some recent experiments~\cite{Xiaoxue_2014_JCIS},
implying that the measured pH deviations might be negligible.  It is
important to note, however, that the next generation of
capsule-based biosensors will rely on higher spatial resolution of
ion densities for practical applications, such as the diagnosis of
cancer~\cite{Weidgans_2006_Science}, and that the magnitude of pH
deviations predicted by our modeling may become correspondingly
significant.

The third issue on which we wish to elaborate is the microcapsule size.
In this report, the outer shell radius is within the range 45-105 nm
(see Fig.~\ref{fig5}), due to limitations of our numerical methods.
In most recent experiments, however, the microcapsules are a few microns
in size~\cite{Kreft_2007_JMC,Xiaoxue_2014_JCIS,delMercato_2011_ACSNano}.
As we have demonstrated (Fig.~\ref{fig5}), the deviation of ion density
induced by a charged microcapsule remains nearly constant upon varying
the capsule size at fixed charge density.  In aqueous solutions, the charge
of a microcapsule is determined mainly by the degree of PE dissociation.
Most recent experiments, however, synthesized PE microcapsules via the
LBL technique, alternately adsorbing PEs onto oppositely charged templates.
For such complex architectures, it is still unknown how sensitive the
degree of dissociation and charge density may be to varying microcapsule size.
Further experiments and simulations are required to quantitatively relate
the charge density and size of LBL-generated PE microcapsules.

Fourth, the PB theory is applicable only under conditions for which
the microions are sufficient in number to be modeled as continuous fields.
Moreover, the mean-field approximation, which neglects ion correlations,
limits applications of the Poisson-Boltzmann theory to solutions
containing only monovalent microions, such as H$^+$, Na$^+$, and K$^+$.
For multivalent ions such as
Ca$^{2+}$~\cite{Clark_1999_AC}, Mg$^{2+}$~\cite{Edwin_2003_AC},
Zn$^{2+}$~\cite{Sumner_2002_Analyst}, and
Fe$^{3+}$~\cite{Sumner_2005_Analyst}, the deviations induced by charged PE
microcapsules may be more complex, due to interionic correlations.
Further simulations and more sophisticated theories that account for
ion discreteness and correlations among multivalent ions are required 
to address these issues.

Finally, we emphasize the great practical importance of microcapsule-based
biosensors, which provide powerful tools for detection of ions in small volumes
and diagnosis of cancer and other
diseases~\cite{Weidgans_2006_Science,RiveraGil_2008_ACSNano,Xie_2011_ACR}.
In aqueous solutions, charged microcapsules are known to redistribute
neighboring ions, inducing deviations of ion densities and local pH.
The results of our theoretical modeling verify the significance of these deviations,
which will become increasingly significant with advances in spatial resolution
of next-generation biosensors, and can guide and facilitate further explorations.

\section{Conclusions}\label{conclusions}
In summary, by implementing the nonlinear Poisson-Boltzmann theory in a cell model,
we have theoretically demonstrated that charged polyelectrolyte microcapsules
can induce deviations of ion concentrations inside their cavities.
Our results show that the capsule charge density and the capsule and salt
concentrations contribute to deviations of ion densities, such as pH, from
their bulk values. Our findings are especially relevant for the design of
microcapsules that encapsulate fluorescent dyes to serve as ionic biosensors
for diagnostic purposes. The theoretical framework developed here can be easily
extended to further investigate ionic strength deviations induced by charged
semipermeable microcapsules.

\vspace*{1cm}
\noindent{\bf \large Acknowledgments} \\[1ex]
This work was supported by the National Science Foundation under
Grant No. DMR-1106331.  We thank Andrew B.~Croll and Damith Rozairo
for discussions.

\balance

\bibliographystyle{rsc}

\begin{mcitethebibliography}{57}
\providecommand*{\natexlab}[1]{#1}
\providecommand*{\mciteSetBstSublistMode}[1]{}
\providecommand*{\mciteSetBstMaxWidthForm}[2]{}
\providecommand*{\mciteBstWouldAddEndPuncttrue}
  {\def\EndOfBibitem{\unskip.}}
\providecommand*{\mciteBstWouldAddEndPunctfalse}
  {\let\EndOfBibitem\relax}
\providecommand*{\mciteSetBstMidEndSepPunct}[3]{}
\providecommand*{\mciteSetBstSublistLabelBeginEnd}[3]{}
\providecommand*{\EndOfBibitem}{}
\mciteSetBstSublistMode{f}
\mciteSetBstMaxWidthForm{subitem}
{(\emph{\alph{mcitesubitemcount}})}
\mciteSetBstSublistLabelBeginEnd{\mcitemaxwidthsubitemform\space}
{\relax}{\relax}

\bibitem[Borisov \emph{et~al.}(2010)Borisov, Mayr, Mistlberger, and
  Klimant]{Borisov_2010_AFR}
S.~M. Borisov, T.~Mayr, G.~Mistlberger and I.~Klimant, \emph{Advanced
  Fluorescence Reporters in Chemistry and Biology II: Molecular Constructions,
  Polymers and Nanoparticles}, 2010, vol.~09, pp. 193--228\relax
\mciteBstWouldAddEndPuncttrue
\mciteSetBstMidEndSepPunct{\mcitedefaultmidpunct}
{\mcitedefaultendpunct}{\mcitedefaultseppunct}\relax
\EndOfBibitem
\bibitem[del Mercato \emph{et~al.}(2010)del Mercato, Rivera-Gil, Abbasi, Ochs,
  Ganas, Zins, Soennichsen, and Parak]{delMercato_2010_Nanoscale}
L.~L. del Mercato, P.~Rivera-Gil, A.~Z. Abbasi, M.~Ochs, C.~Ganas, I.~Zins,
  C.~Soennichsen and W.~J. Parak, \emph{Soft Matter}, 2010, \textbf{2},
  458--467\relax
\mciteBstWouldAddEndPuncttrue
\mciteSetBstMidEndSepPunct{\mcitedefaultmidpunct}
{\mcitedefaultendpunct}{\mcitedefaultseppunct}\relax
\EndOfBibitem
\bibitem[De~Koker \emph{et~al.}(2012)De~Koker, Hoogenboom, and
  De~Geest]{DeKoker_2012_CSR}
S.~De~Koker, R.~Hoogenboom and B.~G. De~Geest, \emph{Chem. Soc. Rev.}, 2012,
  \textbf{41}, 2867--2884\relax
\mciteBstWouldAddEndPuncttrue
\mciteSetBstMidEndSepPunct{\mcitedefaultmidpunct}
{\mcitedefaultendpunct}{\mcitedefaultseppunct}\relax
\EndOfBibitem
\bibitem[De~Koker \emph{et~al.}(2011)De~Koker, De~Cock, Rivera-Gil, Parak,
  Velty, Vervaet, Remon, Grooten, and De~Geest]{Koker_2011_ADDR}
S.~De~Koker, L.~J. De~Cock, P.~Rivera-Gil, W.~J. Parak, R.~A. Velty,
  C.~Vervaet, J.~P. Remon, J.~Grooten and B.~G. De~Geest, \emph{Adv. Drug
  Delivery Rev.}, 2011, \textbf{63}, 748--761\relax
\mciteBstWouldAddEndPuncttrue
\mciteSetBstMidEndSepPunct{\mcitedefaultmidpunct}
{\mcitedefaultendpunct}{\mcitedefaultseppunct}\relax
\EndOfBibitem
\bibitem[De~Geest \emph{et~al.}(2009)De~Geest, De~Koker, Sukhorukov, Kreft,
  Parak, Skirtach, Demeester, De~Smedt, and Hennink]{DeGeest_2009_Softmatter}
B.~G. De~Geest, S.~De~Koker, G.~B. Sukhorukov, O.~Kreft, W.~J. Parak, A.~G.
  Skirtach, J.~Demeester, S.~C. De~Smedt and W.~E. Hennink, \emph{Soft Matter},
  2009, \textbf{5}, 282--291\relax
\mciteBstWouldAddEndPuncttrue
\mciteSetBstMidEndSepPunct{\mcitedefaultmidpunct}
{\mcitedefaultendpunct}{\mcitedefaultseppunct}\relax
\EndOfBibitem
\bibitem[Tong \emph{et~al.}(2012)Tong, Song, and Gao]{Weijun_2012_CSR}
W.~Tong, X.~Song and C.~Gao, \emph{Chem. Soc. Rev.}, 2012, \textbf{41},
  6103--6124\relax
\mciteBstWouldAddEndPuncttrue
\mciteSetBstMidEndSepPunct{\mcitedefaultmidpunct}
{\mcitedefaultendpunct}{\mcitedefaultseppunct}\relax
\EndOfBibitem
\bibitem[Adamczak \emph{et~al.}(2012)Adamczak, Hoel, Gaudernack, Barbasz,
  Szczepanowicz, and Warszynski]{Adamczak_2012_CSB}
M.~Adamczak, H.~J. Hoel, G.~Gaudernack, J.~Barbasz, K.~Szczepanowicz and
  P.~Warszynski, \emph{Colloid Surf. B-Biointerfaces}, 2012, \textbf{90},
  211--216\relax
\mciteBstWouldAddEndPuncttrue
\mciteSetBstMidEndSepPunct{\mcitedefaultmidpunct}
{\mcitedefaultendpunct}{\mcitedefaultseppunct}\relax
\EndOfBibitem
\bibitem[Peteiro-Cattelle \emph{et~al.}(2009)Peteiro-Cattelle,
  Rodriguez-Pedreira, Zhang, Rivera~Gil, del Mercato, and
  Parak]{Peteiro_2009_Nanomedicine}
J.~Peteiro-Cattelle, M.~Rodriguez-Pedreira, F.~Zhang, P.~Rivera~Gil, L.~L. del
  Mercato and W.~J. Parak, \emph{Nanomedicine}, 2009, \textbf{4},
  967--979\relax
\mciteBstWouldAddEndPuncttrue
\mciteSetBstMidEndSepPunct{\mcitedefaultmidpunct}
{\mcitedefaultendpunct}{\mcitedefaultseppunct}\relax
\EndOfBibitem
\bibitem[Caruso \emph{et~al.}(1998)Caruso, Caruso, and
  M{\"o}hwald]{Caruso_1998_Science}
F.~Caruso, R.~A. Caruso and H.~M{\"o}hwald, \emph{Science}, 1998, \textbf{282},
  1111--1114\relax
\mciteBstWouldAddEndPuncttrue
\mciteSetBstMidEndSepPunct{\mcitedefaultmidpunct}
{\mcitedefaultendpunct}{\mcitedefaultseppunct}\relax
\EndOfBibitem
\bibitem[Kida \emph{et~al.}(2006)Kida, Mouri, and Akashi]{Kida_2006_Angew}
T.~Kida, M.~Mouri and M.~Akashi, \emph{Angew. Chem. Int. Ed.}, 2006,
  \textbf{45}, 7534--7536\relax
\mciteBstWouldAddEndPuncttrue
\mciteSetBstMidEndSepPunct{\mcitedefaultmidpunct}
{\mcitedefaultendpunct}{\mcitedefaultseppunct}\relax
\EndOfBibitem
\bibitem[Such \emph{et~al.}(2011)Such, Johnston, and Caruso]{Such_2011_CSR}
G.~K. Such, A.~P.~R. Johnston and F.~Caruso, \emph{Chem. Soc. Rev.}, 2011,
  \textbf{40}, 19--29\relax
\mciteBstWouldAddEndPuncttrue
\mciteSetBstMidEndSepPunct{\mcitedefaultmidpunct}
{\mcitedefaultendpunct}{\mcitedefaultseppunct}\relax
\EndOfBibitem
\bibitem[Kharlampieva and Sukhishvili(2006)]{Kharlampieva_2006_PolymRev}
E.~Kharlampieva and S.~A. Sukhishvili, \emph{Polym. Rev.}, 2006, \textbf{46},
  377--395\relax
\mciteBstWouldAddEndPuncttrue
\mciteSetBstMidEndSepPunct{\mcitedefaultmidpunct}
{\mcitedefaultendpunct}{\mcitedefaultseppunct}\relax
\EndOfBibitem
\bibitem[Wang \emph{et~al.}(2008)Wang, Feng, and Gao]{Wang_2008_CM}
Z.~Wang, Z.~Feng and C.~Gao, \emph{Chem. Mater.}, 2008, \textbf{20},
  4194--4199\relax
\mciteBstWouldAddEndPuncttrue
\mciteSetBstMidEndSepPunct{\mcitedefaultmidpunct}
{\mcitedefaultendpunct}{\mcitedefaultseppunct}\relax
\EndOfBibitem
\bibitem[Zhang \emph{et~al.}(2003)Zhang, Yang, Guan, Cao, and
  Xu]{Zhang_2003_MA}
Y.~J. Zhang, S.~G. Yang, Y.~Guan, W.~X. Cao and J.~Xu, \emph{Macromolecules},
  2003, \textbf{36}, 4238--4240\relax
\mciteBstWouldAddEndPuncttrue
\mciteSetBstMidEndSepPunct{\mcitedefaultmidpunct}
{\mcitedefaultendpunct}{\mcitedefaultseppunct}\relax
\EndOfBibitem
\bibitem[Duan \emph{et~al.}(2007)Duan, He, Yan, Cui, Wang, and
  Li]{Duan_2007_BBRC}
L.~Duan, Q.~He, X.~Yan, Y.~Cui, K.~Wang and J.~Li, \emph{Biochem. Biophys. Res.
  Commun.}, 2007, \textbf{354}, 357--362\relax
\mciteBstWouldAddEndPuncttrue
\mciteSetBstMidEndSepPunct{\mcitedefaultmidpunct}
{\mcitedefaultendpunct}{\mcitedefaultseppunct}\relax
\EndOfBibitem
\bibitem[Johnston \emph{et~al.}(2005)Johnston, Read, and
  Caruso]{Johnston_2005_NanoLett}
A.~P.~R. Johnston, E.~S. Read and F.~Caruso, \emph{Nano Lett.}, 2005,
  \textbf{5}, 953--956\relax
\mciteBstWouldAddEndPuncttrue
\mciteSetBstMidEndSepPunct{\mcitedefaultmidpunct}
{\mcitedefaultendpunct}{\mcitedefaultseppunct}\relax
\EndOfBibitem
\bibitem[Rivera-Gil \emph{et~al.}(2009)Rivera-Gil, De~Koker, De~Geest, and
  Parak]{RiveraGil_2009_NL}
P.~Rivera-Gil, S.~De~Koker, B.~G. De~Geest and W.~J. Parak, \emph{Nano Lett.},
  2009, \textbf{9}, 4398--4402\relax
\mciteBstWouldAddEndPuncttrue
\mciteSetBstMidEndSepPunct{\mcitedefaultmidpunct}
{\mcitedefaultendpunct}{\mcitedefaultseppunct}\relax
\EndOfBibitem
\bibitem[Shen \emph{et~al.}(2011)Shen, Lu, Tian, and Zhu]{Shen_2011_MA}
L.~Shen, X.~Lu, H.~Tian and W.~Zhu, \emph{Macromol.}, 2011, \textbf{44},
  5612--5618\relax
\mciteBstWouldAddEndPuncttrue
\mciteSetBstMidEndSepPunct{\mcitedefaultmidpunct}
{\mcitedefaultendpunct}{\mcitedefaultseppunct}\relax
\EndOfBibitem
\bibitem[Reibetanz \emph{et~al.}(2011)Reibetanz, Chen, Mutukumaraswamy, Liaw,
  Oh, Donath, and Neu]{Reibetanz_2011_JBS}
U.~Reibetanz, M.~H.~A. Chen, S.~Mutukumaraswamy, Z.~Y. Liaw, B.~H.~L. Oh,
  E.~Donath and B.~Neu, \emph{J. Biomater. Sci.}, 2011, \textbf{22},
  1845--1859\relax
\mciteBstWouldAddEndPuncttrue
\mciteSetBstMidEndSepPunct{\mcitedefaultmidpunct}
{\mcitedefaultendpunct}{\mcitedefaultseppunct}\relax
\EndOfBibitem
\bibitem[Kazakova \emph{et~al.}(2011)Kazakova, Shabarchina, and
  Sukhorukov]{Kazakova_2011_PCCP}
L.~I. Kazakova, L.~I. Shabarchina and G.~B. Sukhorukov, \emph{Phys. Chem. Chem.
  Phys.}, 2011, \textbf{13}, 11110--11117\relax
\mciteBstWouldAddEndPuncttrue
\mciteSetBstMidEndSepPunct{\mcitedefaultmidpunct}
{\mcitedefaultendpunct}{\mcitedefaultseppunct}\relax
\EndOfBibitem
\bibitem[Kreft \emph{et~al.}(2007)Kreft, Javier, Sukhorukov, and
  Parak]{Kreft_2007_JMC}
O.~Kreft, A.~M. Javier, G.~B. Sukhorukov and W.~J. Parak, \emph{J. Mater.
  Chem.}, 2007, \textbf{17}, 4471--4476\relax
\mciteBstWouldAddEndPuncttrue
\mciteSetBstMidEndSepPunct{\mcitedefaultmidpunct}
{\mcitedefaultendpunct}{\mcitedefaultseppunct}\relax
\EndOfBibitem
\bibitem[McShane and Ritter(2010)]{McShane_2010_JMC}
M.~McShane and D.~Ritter, \emph{J. Mater. Chem.}, 2010, \textbf{20},
  8189--8193\relax
\mciteBstWouldAddEndPuncttrue
\mciteSetBstMidEndSepPunct{\mcitedefaultmidpunct}
{\mcitedefaultendpunct}{\mcitedefaultseppunct}\relax
\EndOfBibitem
\bibitem[Staedler \emph{et~al.}(2009)Staedler, Price, Chandrawati, Hosta-Rigau,
  Zelikin, and Caruso]{Staedler_2009_Nano}
B.~Staedler, A.~D. Price, R.~Chandrawati, L.~Hosta-Rigau, A.~N. Zelikin and
  F.~Caruso, \emph{Nanoscale}, 2009, \textbf{1}, 68--73\relax
\mciteBstWouldAddEndPuncttrue
\mciteSetBstMidEndSepPunct{\mcitedefaultmidpunct}
{\mcitedefaultendpunct}{\mcitedefaultseppunct}\relax
\EndOfBibitem
\bibitem[Shchukin and Sukhorukov(2004)]{Shchukin_2004_AM}
D.~G. Shchukin and G.~B. Sukhorukov, \emph{Adv. Mater.}, 2004, \textbf{16},
  671--682\relax
\mciteBstWouldAddEndPuncttrue
\mciteSetBstMidEndSepPunct{\mcitedefaultmidpunct}
{\mcitedefaultendpunct}{\mcitedefaultseppunct}\relax
\EndOfBibitem
\bibitem[Kuwana \emph{et~al.}(2004)Kuwana, Liang, and
  Sevick-Muraca]{Kuwana_2004_BP}
E.~Kuwana, F.~Liang and E.~M. Sevick-Muraca, \emph{Biotechnol. Prog.}, 2004,
  \textbf{20}, 1561--1566\relax
\mciteBstWouldAddEndPuncttrue
\mciteSetBstMidEndSepPunct{\mcitedefaultmidpunct}
{\mcitedefaultendpunct}{\mcitedefaultseppunct}\relax
\EndOfBibitem
\bibitem[Reibetanz \emph{et~al.}(2010)Reibetanz, Chen, Mutukumaraswamy, Liaw,
  Oh, Venkatraman, Donath, and Neu]{Reibetanz_2010_BioM}
U.~Reibetanz, M.~H.~A. Chen, S.~Mutukumaraswamy, Z.~Y. Liaw, B.~H.~L. Oh,
  S.~Venkatraman, E.~Donath and B.~Neu, \emph{Biomacromolecules}, 2010,
  \textbf{11}, 1779--1784\relax
\mciteBstWouldAddEndPuncttrue
\mciteSetBstMidEndSepPunct{\mcitedefaultmidpunct}
{\mcitedefaultendpunct}{\mcitedefaultseppunct}\relax
\EndOfBibitem
\bibitem[del Mercato \emph{et~al.}(2011)del Mercato, Abbasi, and
  Parak]{deMercato_2011_Small}
L.~L. del Mercato, A.~Z. Abbasi and W.~J. Parak, \emph{Small}, 2011,
  \textbf{7}, 351--363\relax
\mciteBstWouldAddEndPuncttrue
\mciteSetBstMidEndSepPunct{\mcitedefaultmidpunct}
{\mcitedefaultendpunct}{\mcitedefaultseppunct}\relax
\EndOfBibitem
\bibitem[del Mercato \emph{et~al.}(2011)del Mercato, Abbasi, Ochs, and
  Parak]{delMercato_2011_ACSNano}
L.~L. del Mercato, A.~Z. Abbasi, M.~Ochs and W.~J. Parak, \emph{ACS Nano},
  2011, \textbf{5}, 9668--9674\relax
\mciteBstWouldAddEndPuncttrue
\mciteSetBstMidEndSepPunct{\mcitedefaultmidpunct}
{\mcitedefaultendpunct}{\mcitedefaultseppunct}\relax
\EndOfBibitem
\bibitem[Song \emph{et~al.}(2014)Song, Li, Tong, and Gao]{Xiaoxue_2014_JCIS}
X.~Song, H.~Li, W.~Tong and C.~Gao, \emph{J. Colloid Interface Sci.}, 2014,
  \textbf{416}, 252--257\relax
\mciteBstWouldAddEndPuncttrue
\mciteSetBstMidEndSepPunct{\mcitedefaultmidpunct}
{\mcitedefaultendpunct}{\mcitedefaultseppunct}\relax
\EndOfBibitem
\bibitem[Kazakova \emph{et~al.}(2013)Kazakova, Shabarchina, Anastasova, Pavlov,
  Vadgama, Skirtach, and Sukhorukov]{Kazakova_2013_ABC}
L.~I. Kazakova, L.~I. Shabarchina, S.~Anastasova, A.~M. Pavlov, P.~Vadgama,
  A.~G. Skirtach and G.~B. Sukhorukov, \emph{Anal. Bioanal. Chem.}, 2013,
  \textbf{405}, 1559--1568\relax
\mciteBstWouldAddEndPuncttrue
\mciteSetBstMidEndSepPunct{\mcitedefaultmidpunct}
{\mcitedefaultendpunct}{\mcitedefaultseppunct}\relax
\EndOfBibitem
\bibitem[Sun \emph{et~al.}(2012)Sun, Benjaminsen, Almdal, and
  Andresen]{Sun_2012_BC}
H.~Sun, R.~V. Benjaminsen, K.~Almdal and T.~L. Andresen, \emph{Bioconjugate
  Chem.}, 2012, \textbf{23}, 2247--2255\relax
\mciteBstWouldAddEndPuncttrue
\mciteSetBstMidEndSepPunct{\mcitedefaultmidpunct}
{\mcitedefaultendpunct}{\mcitedefaultseppunct}\relax
\EndOfBibitem
\bibitem[Jimenez~de Aberasturi \emph{et~al.}(2012)Jimenez~de Aberasturi,
  Montenegro, Ruiz~de Larramendi, Rojo, Klar, Alvarez-Puebla, Liz-Marzan, and
  Parak]{Dorleta_2012_CM}
D.~Jimenez~de Aberasturi, J.-M. Montenegro, I.~Ruiz~de Larramendi, T.~Rojo,
  T.~A. Klar, R.~Alvarez-Puebla, L.~M. Liz-Marzan and W.~J. Parak, \emph{Chem.
  Mater.}, 2012, \textbf{24}, 738--745\relax
\mciteBstWouldAddEndPuncttrue
\mciteSetBstMidEndSepPunct{\mcitedefaultmidpunct}
{\mcitedefaultendpunct}{\mcitedefaultseppunct}\relax
\EndOfBibitem
\bibitem[Rivera-Gil \emph{et~al.}(2012)Rivera-Gil, Nazarenus, Ashraf, and
  Parak]{RiveraGil_2012_Small}
P.~Rivera-Gil, M.~Nazarenus, S.~Ashraf and W.~J. Parak, \emph{Small}, 2012,
  \textbf{8}, 943--948\relax
\mciteBstWouldAddEndPuncttrue
\mciteSetBstMidEndSepPunct{\mcitedefaultmidpunct}
{\mcitedefaultendpunct}{\mcitedefaultseppunct}\relax
\EndOfBibitem
\bibitem[Lee \emph{et~al.}(2011)Lee, Tiwari, and Raghavan]{Lee_2011_Softmatter}
H.-Y. Lee, K.~R. Tiwari and S.~R. Raghavan, \emph{Soft Matter}, 2011,
  \textbf{7}, 3273--3276\relax
\mciteBstWouldAddEndPuncttrue
\mciteSetBstMidEndSepPunct{\mcitedefaultmidpunct}
{\mcitedefaultendpunct}{\mcitedefaultseppunct}\relax
\EndOfBibitem
\bibitem[Patel \emph{et~al.}(2013)Patel, Remijn, Cabero, Heussen, ten Hoorn,
  and Velikov]{Patel_2013_AFM}
A.~R. Patel, C.~Remijn, A.-i.~M. Cabero, P.~C.~M. Heussen, J.~W. M.~S. ten
  Hoorn and K.~P. Velikov, \emph{Adv. Funct. Mater.}, 2013, \textbf{23},
  4710--4718\relax
\mciteBstWouldAddEndPuncttrue
\mciteSetBstMidEndSepPunct{\mcitedefaultmidpunct}
{\mcitedefaultendpunct}{\mcitedefaultseppunct}\relax
\EndOfBibitem
\bibitem[Janata(1987)]{Janata_1987_AC}
J.~Janata, \emph{Anal. Chem.}, 1987, \textbf{59}, 1351--1356\relax
\mciteBstWouldAddEndPuncttrue
\mciteSetBstMidEndSepPunct{\mcitedefaultmidpunct}
{\mcitedefaultendpunct}{\mcitedefaultseppunct}\relax
\EndOfBibitem
\bibitem[Janata(1992)]{Janata_1992_AC}
J.~Janata, \emph{Anal. Chem.}, 1992, \textbf{64}, 921A--927A\relax
\mciteBstWouldAddEndPuncttrue
\mciteSetBstMidEndSepPunct{\mcitedefaultmidpunct}
{\mcitedefaultendpunct}{\mcitedefaultseppunct}\relax
\EndOfBibitem
\bibitem[Bostrom \emph{et~al.}(2002)Bostrom, Williams, and
  Ninham]{Bostrom_2002_La}
M.~Bostrom, D.~R.~M. Williams and B.~W. Ninham, \emph{Langmuir}, 2002,
  \textbf{18}, 8609--8615\relax
\mciteBstWouldAddEndPuncttrue
\mciteSetBstMidEndSepPunct{\mcitedefaultmidpunct}
{\mcitedefaultendpunct}{\mcitedefaultseppunct}\relax
\EndOfBibitem
\bibitem[Zhang \emph{et~al.}(2010)Zhang, Ali, Amin, Feltz, Oheim, and
  Parak]{Zhang_2011_CPC}
F.~Zhang, Z.~Ali, F.~Amin, A.~Feltz, M.~Oheim and W.~J. Parak, \emph{Chem.
  Phys. Chem.}, 2010, \textbf{11}, 730--735\relax
\mciteBstWouldAddEndPuncttrue
\mciteSetBstMidEndSepPunct{\mcitedefaultmidpunct}
{\mcitedefaultendpunct}{\mcitedefaultseppunct}\relax
\EndOfBibitem
\bibitem[Weissleder(2006)]{Weidgans_2006_Science}
R.~Weissleder, \emph{Science}, 2006, \textbf{312}, 1168--1171\relax
\mciteBstWouldAddEndPuncttrue
\mciteSetBstMidEndSepPunct{\mcitedefaultmidpunct}
{\mcitedefaultendpunct}{\mcitedefaultseppunct}\relax
\EndOfBibitem
\bibitem[Rivera~Gil and Parak(2008)]{RiveraGil_2008_ACSNano}
P.~Rivera~Gil and W.~J. Parak, \emph{ACS Nano}, 2008, \textbf{2},
  2200--2205\relax
\mciteBstWouldAddEndPuncttrue
\mciteSetBstMidEndSepPunct{\mcitedefaultmidpunct}
{\mcitedefaultendpunct}{\mcitedefaultseppunct}\relax
\EndOfBibitem
\bibitem[Xie \emph{et~al.}(2011)Xie, Liu, Eden, Ai, and Chen]{Xie_2011_ACR}
J.~Xie, G.~Liu, H.~S. Eden, H.~Ai and X.~Chen, \emph{Acc. Chem. Res.}, 2011,
  \textbf{44}, 883--892\relax
\mciteBstWouldAddEndPuncttrue
\mciteSetBstMidEndSepPunct{\mcitedefaultmidpunct}
{\mcitedefaultendpunct}{\mcitedefaultseppunct}\relax
\EndOfBibitem
\bibitem[Sukhorukov \emph{et~al.}(1999)Sukhorukov, Brumen, Donath, and
  M{\"o}hwald]{Sukhorukov_1999_JPCB}
G.~B. Sukhorukov, M.~Brumen, E.~Donath and H.~M{\"o}hwald, \emph{J. Phys. Chem.
  B}, 1999, \textbf{103}, 6434--6440\relax
\mciteBstWouldAddEndPuncttrue
\mciteSetBstMidEndSepPunct{\mcitedefaultmidpunct}
{\mcitedefaultendpunct}{\mcitedefaultseppunct}\relax
\EndOfBibitem
\bibitem[Halo{\v z}an \emph{et~al.}(2005)Halo{\v z}an, D{\'e}jugnat, Brumen,
  and Sukhorukov]{David_2005_JCIM}
D.~Halo{\v z}an, C.~D{\'e}jugnat, M.~Brumen and G.~B. Sukhorukov, \emph{J.
  Chem. Inf. Model.}, 2005, \textbf{45}, 1589--1592\relax
\mciteBstWouldAddEndPuncttrue
\mciteSetBstMidEndSepPunct{\mcitedefaultmidpunct}
{\mcitedefaultendpunct}{\mcitedefaultseppunct}\relax
\EndOfBibitem
\bibitem[Halo{\v z}an \emph{et~al.}(2007)Halo{\v z}an, Sukhorukov, Brumen,
  Donath, and M{\"o}hwald]{David_2007_ACS}
D.~Halo{\v z}an, G.~B. Sukhorukov, M.~Brumen, E.~Donath and H.~M{\"o}hwald,
  \emph{Acta Chim. Slov.}, 2007, \textbf{54}, 598--604\relax
\mciteBstWouldAddEndPuncttrue
\mciteSetBstMidEndSepPunct{\mcitedefaultmidpunct}
{\mcitedefaultendpunct}{\mcitedefaultseppunct}\relax
\EndOfBibitem
\bibitem[Marcus(1955)]{Marcus_1955_JCP}
R.~A. Marcus, \emph{J. Chem. Phys.}, 1955, \textbf{23}, 1057--1068\relax
\mciteBstWouldAddEndPuncttrue
\mciteSetBstMidEndSepPunct{\mcitedefaultmidpunct}
{\mcitedefaultendpunct}{\mcitedefaultseppunct}\relax
\EndOfBibitem
\bibitem[Parthasarathy and Klingenberg(1996)]{Parthasarathy_1996_MSE}
M.~Parthasarathy and D.~J. Klingenberg, \emph{Mater. Sci. Eng.}, 1996,
  \textbf{17}, 57--103\relax
\mciteBstWouldAddEndPuncttrue
\mciteSetBstMidEndSepPunct{\mcitedefaultmidpunct}
{\mcitedefaultendpunct}{\mcitedefaultseppunct}\relax
\EndOfBibitem
\bibitem[Mohanty \emph{et~al.}(2012)Mohanty, Yethiraj, and
  Schurtenberger]{Mohanty_2012_SM}
P.~S. Mohanty, A.~Yethiraj and P.~Schurtenberger, \emph{Soft Matter}, 2012,
  \textbf{8}, 10819--10822\relax
\mciteBstWouldAddEndPuncttrue
\mciteSetBstMidEndSepPunct{\mcitedefaultmidpunct}
{\mcitedefaultendpunct}{\mcitedefaultseppunct}\relax
\EndOfBibitem
\bibitem[Wennerstr{\"o}m \emph{et~al.}(1982)Wennerstr{\"o}m, J{\"o}nsson, and
  Linse]{Hakan_1982_JCP}
H.~Wennerstr{\"o}m, B.~J{\"o}nsson and P.~Linse, \emph{J. Chem. Phys.}, 1982,
  \textbf{76}, 4665--4670\relax
\mciteBstWouldAddEndPuncttrue
\mciteSetBstMidEndSepPunct{\mcitedefaultmidpunct}
{\mcitedefaultendpunct}{\mcitedefaultseppunct}\relax
\EndOfBibitem
\bibitem[Denton(2003)]{Denton_2003_PRE}
A.~R. Denton, \emph{Phys. Rev. E}, 2003, \textbf{67}, 011804--1--10\relax
\mciteBstWouldAddEndPuncttrue
\mciteSetBstMidEndSepPunct{\mcitedefaultmidpunct}
{\mcitedefaultendpunct}{\mcitedefaultseppunct}\relax
\EndOfBibitem
\bibitem[Denton(2010)]{Denton_2010_JPCM}
A.~R. Denton, \emph{J. Phys.: Condens. Matter}, 2010, \textbf{22},
  364108--1--8\relax
\mciteBstWouldAddEndPuncttrue
\mciteSetBstMidEndSepPunct{\mcitedefaultmidpunct}
{\mcitedefaultendpunct}{\mcitedefaultseppunct}\relax
\EndOfBibitem
\bibitem[Press \emph{et~al.}(2007)Press, Teukolsky, Vetterling, and
  Flannery]{Numerical_Recipes_2007}
W.~H. Press, S.~A. Teukolsky, W.~T. Vetterling and B.~P. Flannery,
  \emph{Numerical Recipes: The Art of Scientific Computing}, Cambridge
  University Press, New York, 3rd edn, 2007\relax
\mciteBstWouldAddEndPuncttrue
\mciteSetBstMidEndSepPunct{\mcitedefaultmidpunct}
{\mcitedefaultendpunct}{\mcitedefaultseppunct}\relax
\EndOfBibitem
\bibitem[Weidgans \emph{et~al.}(2004)Weidgans, Krause, Klimant, and
  Wolfbeis]{Weidgans_2004_Analyst}
B.~M. Weidgans, C.~Krause, I.~Klimant and O.~S. Wolfbeis, \emph{Analyst}, 2004,
  \textbf{129}, 645--650\relax
\mciteBstWouldAddEndPuncttrue
\mciteSetBstMidEndSepPunct{\mcitedefaultmidpunct}
{\mcitedefaultendpunct}{\mcitedefaultseppunct}\relax
\EndOfBibitem
\bibitem[Clark \emph{et~al.}(1999)Clark, Kopelman, Tjalkens, and
  Philbert]{Clark_1999_AC}
H.~A. Clark, R.~Kopelman, R.~Tjalkens and M.~A. Philbert, \emph{Anal. Chem.},
  1999, \textbf{71}, 4837--4843\relax
\mciteBstWouldAddEndPuncttrue
\mciteSetBstMidEndSepPunct{\mcitedefaultmidpunct}
{\mcitedefaultendpunct}{\mcitedefaultseppunct}\relax
\EndOfBibitem
\bibitem[Park \emph{et~al.}(2003)Park, Brasuel, Behrend, Philbert, and
  Kopelman]{Edwin_2003_AC}
E.~J. Park, M.~Brasuel, C.~Behrend, M.~A. Philbert and R.~Kopelman, \emph{Anal.
  Chem.}, 2003, \textbf{75}, 3784--3791\relax
\mciteBstWouldAddEndPuncttrue
\mciteSetBstMidEndSepPunct{\mcitedefaultmidpunct}
{\mcitedefaultendpunct}{\mcitedefaultseppunct}\relax
\EndOfBibitem
\bibitem[Sumner \emph{et~al.}(2002)Sumner, Aylott, Monson, and
  Kopelman]{Sumner_2002_Analyst}
J.~P. Sumner, J.~W. Aylott, E.~Monson and R.~Kopelman, \emph{Analyst}, 2002,
  \textbf{127}, 11--16\relax
\mciteBstWouldAddEndPuncttrue
\mciteSetBstMidEndSepPunct{\mcitedefaultmidpunct}
{\mcitedefaultendpunct}{\mcitedefaultseppunct}\relax
\EndOfBibitem
\bibitem[Sumner and Kopelman(2005)]{Sumner_2005_Analyst}
J.~P. Sumner and R.~Kopelman, \emph{Analyst}, 2005, \textbf{130},
  528--533\relax
\mciteBstWouldAddEndPuncttrue
\mciteSetBstMidEndSepPunct{\mcitedefaultmidpunct}
{\mcitedefaultendpunct}{\mcitedefaultseppunct}\relax
\EndOfBibitem
\end{mcitethebibliography}


\providecommand*{\mcitethebibliography}{\thebibliography}
\csname @ifundefined\endcsname{endmcitethebibliography}
{\let\endmcitethebibliography\endthebibliography}{}

\end{document}